\documentclass{jetpl}
\usepackage{amsmath,amsfonts,epsfig,graphics,amssymb}

\usepackage{hyperref}
\hypersetup{
colorlinks=true
}

\lat
\title{Ultra-high energy cosmic ray correlations with Active Galactic Nuclei in the world dataset}
\rtitle{UHECR correlations with AGN \dots}
\sodtitle{Ultra-high energy cosmic ray correlations with Active Galactic Nuclei in the world dataset}

\author{
G.\,I.\,Rubtsov$^{a}$, %\/\thanks{e-mail:grisha@ms2.inr.ac.r},
I.\,I.\,Tkachev$^{a,b}$, and
A.\,D.\,Dolgov$^{b,c,d,e}$
}

\sodauthor{Dolgov, Rubtsov, Tkachev}

\address{$^a$Institute  for Nuclear Research of the Russian Academy of Sciences,
Moscow 117312, Russia\\~\\
$^b$Laboratory of Cosmology and Elementary Particles, Novosibirsk State University, Pirogov street 2, 630090 Novosibirsk, Russia\\~\\
$^c$Dipartimento di Fisica, Universit`a degli Studi di Ferrara,  Polo Scientifico e Tecnologico - Edificio C, Via Saragat 1, 44122 Ferrara, Italy\\~\\
$^d$Istituto Nazionale di Fisica Nucleare, Sezione di Ferrara,
Polo Scientifico e Tecnologico - Edificio C, Via Saragat 1, 44122 Ferrara, Italy
\\~\\
$^e$Institute of Theoretical and Experimental Physics,
Bolshaya Cheremushkinskaya ul. 25, 113259 Moscow, Russia\\~\\
}

\abstract{Pierre Auger collaboration have recently put forward the
  hypothesis that the arrival directions of the highest energy cosmic
  rays correlate with the subset of local active galactic nuclei
  (AGN). We perform a blind test of AGN hypothesis using publicly
  available event sets collected by Yakutsk, AGASA and HiRes
  experiments. The consistency of the procedure requires the event
  energies to be normalized towards the common energy scale. The
  number of correlating events in resulting data-set is 3 of 21 which is consistent with expected random background.}
\PACS{98.70.Sa, 96.50.sb, 96.50.sd}

\newcommand{\be}{\begin{equation}}
\newcommand{\ee}{\end{equation}}

\begin{document}

\maketitle

\textbf{1. Introduction.} 

The question of the origin of ultra-high energy cosmic rays (UHECR) belongs to the list of the most interesting unsolved  problems in particle astrophysics. Many very different models were suggested here, including decaying superheavy dark matter, cosmic strings, gamma-ray bursts, AGN, etc.  Detection of anisotropy of arrival directions of primaries would be a key to the resolution of this puzzle and identification of  the UHECR production mechanism, which is, therefore, of fundamental importance. Establishing the level of anisotropy is also an important step in ironing out the chemical composition of UHECR and measuring parameters of the intergalactic medium such as the strength of magnetic fields and intensity of cosmic background radiation. Search for correlations between cosmic rays (CR) arrival directions and positions on the celestial sphere of prospective astrophysical sources is one of the promising methods to approach these problems, see e.g. \cite{Tinyakov:2001nr}.

Recently, Pierre Auger Observatory has reported
correlation~\cite{Cronin:2007zz} between the highest energy cosmic
rays and population of all known nearby (closer than 75 Mpc) AGNs. The
correlation was observed at an angle of 3.1$^\circ$. In the control
data set, the number of correlating events was 9 out of 13, which
corresponds to about 69\% of events. This analysis has been later
updated by the Auger collaboration using larger dataset of latest
events \cite{:2010zzj}. In the latter, the smaller fraction of UHECR
correlates with the same set of AGNs. Namely, arrival directions of 28
events were found to correlate with AGNs positions, out of 84 events
with $E > 55$ EeV. This corresponds to the fraction of correlating
events equal to 33\%.

This important claim has been put to the test by other CR
collaborations using their own data, namely, by the
HiRes~\cite{Abbasi:2008md}, Yakutsk~\cite{Ivanov:2008it} and the
Telescope Array~\cite{Tinyakov:2011zz}. The results obtained range
from support of correlation claim \cite{Ivanov:2008it} to complete
absence of correlations~\cite{Abbasi:2008md}.

It should be stressed, however, that a verification of such delicate
hypotheses as a claim of correlations, should be done on the basis of
so called "blind test". Namely, all cuts on energy, source catalog,
angular scale of correlations, etc., should be kept strictly at the
same values as in original hypotheses. Requirements of keeping the
original source catalog and unchanged angular scale are easy to
fulfill. On the contrary, the issue with CR catalogs is more
complicated, especially regarding the cut on the energy. The
complication stems from the fact that different experiments have
different energy scales. This is apparent since cosmic ray fluxes
measured by different experiments differ at the same values of nominal
energy, even in the energy regions where statistics is abundant and
there is no doubt that the physical sky is close to isotropic.

Requirement of choosing the same energy cut for all experiments is
extremely important. We are searching for correlation signal with the
sources in the region of Greisen-Zatsepin-Kuzmin \cite{GZK}
suppression. Here the horizon (i.e. the distance to sources from which
CR can reach us) changes exponentially with energy. Small change in
energy leads to dramatic change in horizon and, respectively, in the
subset of sources being probed. The cross-calibration of different
experiments is therefore pressing issue.

Despite of importance, tests of AGN correlation hypothesis using common energy scale for different experiments has not been done in the past. We do it in this paper.

\textbf{2. Energy scales estimate.} 

As it was first shown by V.~Berezinsky~\cite{Berezinsky:2008qh} the
cosmic ray spectra measured by different giant air-shower observatories
become identical at lower energies after simple overall rescaling of energy
\be
  E \rightarrow E \times C,
\ee
 with the constant factor $C$, which
is different for individual experiments.  This opens the way for
cross-calibration of energy scales, c.f. \cite{Gorbunov:2003ra}.
 To handle this issue the
Spectrum Working Group (WG) has been recently formed consisting of members of Auger,
Telescope Array, HiRes and Yakutsk collaborations. Results were
reported at International Symposium on Future Directions in UHECR
Physics, UHECR2012~\cite{uhecr12}. %~\footnote{\url{http://2012.uhecr.org}}.  
Note, that
true, physical, energy scale cannot be determined in this approach. One
may only make sure that quoted values of energy for different
experiments are based on the same energy scale. However, this is sufficient for our
purpose of blind testing the correlation hypothesis. For
definiteness and in order to avoid biases towards one or the other
modern experiments, it was decided to choose common energy scale as a
mean between Auger and Telescope Array energy scales.  
\be
 E_{\rm com}
= \frac{1}{2}(E_{\rm Auger}+E_{\rm TA}).  
\ee

Resulting values of the scaling constant $C$, reported by the WG~\cite{WG}, are reproduced in Table~1. 
\footnote{% WC presentation can be found at 
\url{%
http://indico.cern.ch/contributionDisplay.py?contribId=6&confId=152124}}

\begin{table}[htb]
\begin{center}
\begin{tabular}{|c|c|}
\hline
Experiment & C\\
\hline
Auger      & ~~1.102~~ \\
~Telescope Array~ & 0.906\\
HiRes & 0.911\\
AGASA & 0.652\\
Yakutsk  & 0.561 \\ 
\hline
\end{tabular}
\end{center}
\caption{\label{Tab:exp} Table 1. Values of the energy scaling constant, $C$, for listed experiments.}
\end{table}

\textbf{3. Correlation study with AGNs.} 

For this study we used publicly available catalogs of cosmic ray
events of AGASA~\cite{Hayashida:2000zr}, Yakutsk~\cite{Yakutsk} and
HiRes-stereo~\cite{HiResCatalog}. Telescope Array data are not
publicly available yet.

Auger energy cut of 55 EeV \cite{:2010zzj} corresponds to the cut
$E_{\rm com} > 60.61$ EeV in terms of the common energy scale
adopted. With this energy cut we have 21 events in combined world
catalog. This makes largest to date public event list in the Northern
Hemisphere for testing AGN hypothesis. Out of these 21 events 9 are
from HiRes, 3 from Yakutsk and 9 contributed by AGASA.

\begin{figure}[htb]
\begin{center}
\includegraphics[width=0.75\textwidth]{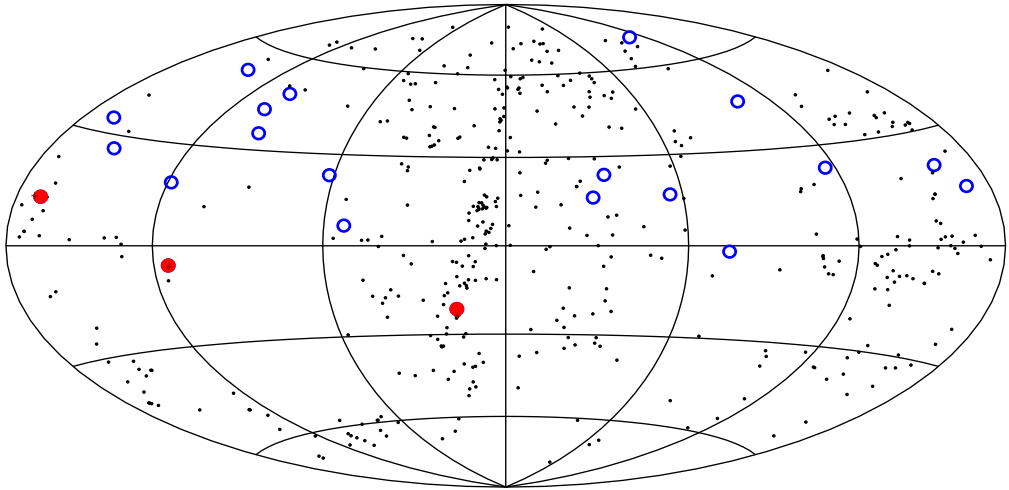}
\end{center}
\caption{Figure 1. Hammer projection of CR arrival directions of Yakutsk, AGASA
  and HiRes-stereo in equatorial coordinate system with $E_{\rm com} > 60.61$ EeV. Correlating and
  non-correlating events are represented by filled red and empty blue
  circles, respectively. AGNs are shown by black dots. \label{Fig:map}}
\end{figure}

Only 3 out of 21 events correlate with AGNs within 3.1 degrees, while
the expected background of random coincidences for the isotropic
distribution of arrival directions equals to 5. Map of arrival
directions in equatorial coordinate system is shown in
Fig.~\ref{Fig:map}. We see that correlations are absent not only with
AGNs as point sources, but, remarkably, CR events are avoiding the
Large Scale Structure. In particular, there are no events from the
Virgo region (represented by the dense concentration of dots near the center of Fig.~1), while there is one event in reasonable vicinity (22 degrees) from
Cen~A despite exposure is low here (cf. Ref.~\cite{Gorbunov:2007ja}).

\newpage

\textbf{4. Conclusions.}  

Common energy scale established by the WG opened up the possibility to
search for correlations using the combined dataset. Based on the
common energy scale we performed a blind test of Pierre Auger AGN
hypothesis using the data of Yakutsk, AGASA and HiRes experiments. The
analysis shown 3 correlating events out of 21, which is compatible
with the expected random background of 5 events. Difference in the fraction
of correlating events in South and North may be attributed to the
existence of peculiar sources in South
(e.g. Cen~A~\cite{Gorbunov:2007ja}) or to statistical fluctuations.

{\bf Acknowledgments.}  We are indebted to P.\,Tinyakov and S.\,Troitsky for useful discussions and criticism. 

This work was supported in part by the Grant of the Government of Russian Federation (11.G34.31.0047) and by the RFBR grants 10-02-01406a, 11-02-01528a, 
12-02-91323-SIGa, by the grants of the President of the Russian Federation NS-5590.2012.2 and MK-1632.2011.2.

%\section*{References}

\end{document}